\documentstyle[aps,prl,twocolumn,epsfig,amsmath,amssymb]{revtex}

\newcommand{\epsfigs}[3]{
\begin{figure}[htb]
\begin{center}
\vskip1.0cm
\resizebox{7cm}{!}{
\includegraphics{#1}}
\caption{\label{#3}#2}
\end{center}
\end{figure}}

\begin{document}
\twocolumn[\hsize\textwidth
\columnwidth\hsize\csname
@twocolumnfalse\endcsname

\title{Monte Carlo study of the elastic interaction in heteropitaxial growth}
\author{B. M. T. Gon\c calves$^\ast$ and J. F. F. Mendes$^\dagger$}
\address{Departamento de F\'\i sica da Universidade do Porto\\
Centro de F\'\i sica do Porto, Rua do Campo Alegre, 687\\
4169-007 Porto, Portugal}
\date{\today }
\maketitle

\begin{abstract}
We have studied the island size distribution and spatial correlation function
of an island growth model under the effect of an  
elastic interaction of the form $1/r^{3}$. The mass 
distribution $P_n(t)$ that was obtained presents a pronounced peak 
that widens with the increase of the total coverage of the system, 
$\theta$. The presence of this peak is an indication of 
the self-organization of the system, since it demonstrates that some sizes are 
more frequent than others.
We have treated exactly the energy of the system using periodic boundary
conditions which were used in the Monte-Carlo simulations.
A discussion about the effect of different factors is presented.
\vskip1cm
\end{abstract}
]

\narrowtext

\section{Introduction}
Epitaxial growth has been the focus of much interest in the past 
years. This interest is derived mainly from the fact that this kind 
of processes have numerous applications \cite{saito96,notzel94,julien93} 
in developing new types of 
devices and materials with some special characteristics.  
Among these processes, there is one type of growth characterized by the 
presence of 
long-range elastic interactions that play a special role \cite{marchenko80}. 
Molecular Beam Epitaxy (MBE) under the effect of elastic interactions 
has been a topic of recent research, with many applications, specially 
in the fabrication of low 
dimensional nanostructures, like quantum dots (QD) and quantum wires 
\cite{notzel94,brune98,pohl99}. One challenging question 
concerning this topic is how, 
and by which mechanism, island organization \cite{floro2000,moisson94} occurs. 
A wide range of 
material/substrate combinations have been observed (e.g., InAS on GaAs).
When one kind of material is deposited over a different one --, usually, with
a different lattice parameter -- it will induce, through this 
structural difference,
a long-range elastic interaction between the deposited 
atoms during surface growth.
The deformation thus obtained originates a strain 
in the substrate that causes different particles to repel each other.
This kind of interaction is supposed to be the mechanism responsible 
of the self-organization observed experimentally.
Many recent studies were developed in order to identify the influence 
of strain on epitaxial and surface morphology during growth 
\cite{mattsson99,Pelya99,brener,alb97,stein,politi2000,saito98}.

Several authors have studied the effect that strain induces on epitaxial 
growth on different types of systems. For instance, the effect of 
elastic strain on the properties of the well know 
Eden model\cite{saito98}, 
and for other versions of a harmonic interaction between the lattice 
atoms\cite{Saito1,Saito2}. Also, the Lennard-Jones 
potential was used to study a similar phenomena \cite{Shindler}. 

In this paper we shall consider that the strain 
induced on the system is due to an repulsive elastic interaction between the 
deposited particles proportional to $1/r^{3}$. 
This type of potential can be derived 
from elasticity theory considerations 
\cite{marchenko80,landau91,hardy67,lau77,andreev81}, when a 
lattice distortion is created (e.g. by cutting out a sphere of the 
bulk and substituting it by a different radius sphere) a field of 
lattice strains is created. It is already known that this 
type of long-range interaction can 
be applied to the absorption of atoms onto a surface but is only valid 
on the case of very thin absorbed clusters (submonolayer regime); 
in more general cases it can be obtained by scaling laws \cite{politi2000}.

\section{Elastic Interaction Potential}

We define the elastic potential to be of the form, $E=Gm_im_j/r^3$,
where $r$ represents the distance between the particles, $m_i$ is 
the ``mass'' of the particle $i$, and $G$ is the coupling constant. 
$G$ usually depends on the elastic properties of the substrate, such 
as the Young modulus, the Poisson ratio and the lattice mismatch.
The coupling constant is given by \cite{Pelya99} as being,
$G=\pi (1-\sigma^2)a^{2}f^{2}/E$,
but other authors define it in different ways \cite{marchenko80,saito98}.
The numerical values of the various forms differ on several orders of 
magnitude. 
We overcome this difficulty by leaving the analythical form of $G$ unspecified 
and determining a reasonable numerical value for it by physical 
considerations.

Since we are trying to study the behavior of a macroscopic material, we 
employed periodic boundary conditions during the simulations. The presence 
of this type of boundary conditions implies that some treatment must be 
given to the energy defined above in order to avoid some 
undesirable ``finite-size effects'' that could originate ``unphysical'' 
results. To accomplish this, we consider an infinite succession of 
replicas of the 
system and calculate the total energy of the infinite system thus obtained. 
The total energy has the contribution of two components, the first being 
the interaction energy between the particle that is currently suffering the 
absorption on the original system and all the other copies of this particle 
that belong to the other systems. This contribution is given by
\begin{equation}
E_1=\sum_{\forall i}\frac{Gm_i^2}{L^3}\sum_{\forall k\neq 0}
\frac{1}{k^3}=\frac{2G\zeta(3)}{L^3}\sum_{\forall i}m_i^2
\end{equation}
where $L$ is the linear dimension of the system and $\zeta(x)$ is 
the Riemman Zeta function. The first sum is performed over all particles 
present in the system at this time.

The second contribution to the total energy is given by the interaction 
between the deposited particle and all particles deposited previously
 in the system. This contribution is expressed as,
\begin{equation}
E_2=\sum_{i<j}\frac{Gm_im_j}{L^3}\sum_{k=-\infty}^\infty\frac{1}{(a_{ij}+k)^3}.
\end{equation}
where the first sum is performed over all pairs of particles, $a_{ij}$ is 
the distance that separates the two particles in the original system divided 
by $L$. Considering that
\begin{equation}
\sum_{k=0}^{\infty}\frac{1}{(k+a)^3}=-\frac{1}{2}\psi^{(2)}(a)
\end{equation} 
where $\psi^{(2)}(a)$ is the second order Polygamma function, and  
after performing some straightforward algebraic manipulations, 
the energy $E_2$ takes the 
following form,
\begin{equation}
E_2=\sum_{i<j}\frac{Gm_im_j}{L^3}\left\{\frac{1}{a_{ij}^3}-\frac{1}{2}
\left(\psi^{(2)}(a_{ij})+\psi^{(2)}(-a_{ij})\right) \right\}.
\end{equation}
The total energy of the system is then given by
\begin{multline}
E_T=\frac{2G\zeta(3)}{L^3}\sum_{i}m_i^2+\\\sum_{i<j}\frac{Gm_im_j}{L^3}
\left\{\frac{1}{a_{ij}^3}-\frac{1}{2}\left(\psi^{(2)}(a_{ij})+\psi^{(2)}
(-a_{ij})\right) \right\}
\end{multline}

During the simulations, and since we are not interested in the absolute value 
of the energy but only in energy differences, we shall only consider 
the ``effective'' value of the energy, that is, the part of the energy 
that is not constant, in order to shorten the CPU time without 
loss of precision in the results.

\section{Simulations}

The model described earlier was implemented in a relatively simple way. 
We consider the substrate to be one dimensional and we shall only consider 
the regime of submonolayer growth.
One site of the system is selected randomly. If that site is occupied, 
the deposition attempt fails and another site is selected. If the 
selected position is empty, three possible situations can occur according 
to the number of nearest neighbor (NN) sites that are occupied.
When only one NN is occupied, the particle adheres irreversibly to 
the preexisting cluster and another deposition attempt is performed. 
If the two NN are occupied, the particle adheres to both clusters, 
coalescing them to become one single cluster with mass conservation. 
Finally, if none of the NN sites is occupied, the 
particle diffuses, due to the repulsive effect of the potential generated 
by the mass distribution present in the system, moving away from the larger 
cluster and becoming closer to the smaller one, until it reaches the local 
minimum of the energy. At each diffusion step, the energy resulting from the 
interaction of the adatom with every other particle present in the system is
calculated.
At this point, the particle begins to diffuse due 
to the effect of the temperature with probability proportional to 
$e^{-\Delta E/k_BT}$, where $E$ is the total energy of the 
system \cite{nota1}. 
During this process, a number $D$ of random 
steps is performed. If during this random walk motion, the 
particle collides with another particle, it aggregates irreversibly 
and another particle is deposited.
This model has several adjustable parameters such as the temperature $T$, the 
number of diffusion steps $D$ and the value of the interaction constant $G$. 
In order to make the simulations behave realistically, these 
parameters must be adjusted and their effect on the final result must 
be studied and well understood. This has been done by varying the 
parameters in order to see the effect that each parameter individually has 
on the final result.
During the experimental study of this type of processes, one usually uses 
temperatures in the interval $300K\leq T \leq 1000K$. To use this 
values during the simulations, one must adjust $G$ in a way as to make the 
factor $\Delta E/k_BT\sim 1$. The typical value of the energy differences 
in this model is of the order $10^{9}$, and considering that $T\sim 10^2$ 
we find that $Gm^2\Delta E_{ef}/(L^3 k_BT)\sim 1$,
and so, $Gm^2/L^3k_B\sim10^{-7}$.

\section{Results}

In this section we will present the results obtained using 
Monte Carlo simulations. To characterize the coarsening dynamics, 
two quantities were sampled and averaged over the initial 
conditions: cluster ``mass'' distribution, 
$P_n(t)$ and the correlation function at equal times $C(r,t)$, defined by
\begin{equation}
\label{corr}
C(r,t) = \langle \rho(r'+r,t)\rho(r',t) \rangle - 
\langle \rho(r',t)\rangle^{2}
\end{equation}
where $\rho(r,t)$ is the site density. The most convincing result yielding 
the self-organization process is the fact that the ``mass'' 
distribution function presents a well defined peak.

\epsfigs{fig1_GM}{Mass distribution variation with the increase of the 
system coverage. Notice that the distribution function widens as the 
coverage increases. In the inset we represent the number of clusters 
present in the system as a function of coverage. It is visible that 
after a certain point, the number of clusters decreases due to the 
occurrence of coalescence phenomena. In all the simulations, $T=500K$ 
and $D=200$ steps.}{fig1_GM}

The shape of the distribution is maintained as one increases the coverage, 
but the height of the function tends to decrease as the width increases. This 
fact was expected to happen, larger values of coverage imply that 
fewer individual clusters are present in the system, but with large sizes. 

There exist two basic process that the system has available in order 
to organize itself. The first, the nucleation of new clusters, is dominant 
in the early stages of the system evolution, when the coverage is small 
and the adatoms never collide. The second, is the coalescence 
of existing clusters. This process becomes dominant as the coverage 
increases, originating larger clusters but in a smaller number. This two 
regimes are clearly seen in the inset of Fig.\ref{fig1_GM}. In the 
beginning, the number of clusters in the system seems to grow almost 
linearly with the coverage. Afterwards, there exists a crossover period 
when the number of clusters is approximately constant, that happens when 
the growth of existing clusters becomes more frequent than the 
nucleation of new ones. Finally, coalescence begins to dominate 
the dynamics and the number of clusters in the system diminishes 
until it becomes one when the coverage gets very large.

The behaviour of the system is described by the island-size 
distribution function, $P_n (\theta)$. Assuming there 
exists a scaling for $P_n (\theta)$ 
one may write, $P_n(\theta)=\lambda f(\lambda^\alpha n,\lambda^\beta\theta)$.
The coverage $\theta$ grows with time, but satisfies, 
$\theta \equiv \sum_{n\ge 1} n P_n(\theta)$. This sum can be approximated by  
an integral, $\theta \approx \int_{0}^{\infty} n P_n(\theta) dn$ resulting 
a relation between the exponents, $\beta=2\alpha-1$. Therefore, one can write,
\begin{equation}
P_n(\theta)=\theta^{1-2\alpha/\beta}\Phi\left(\frac{n}{\theta^{\alpha/\beta}}\right)
\end{equation}
All this results follows from the assumption that there exist only 
one characteristic size in the system, the average island size, 
$S = \sum_{n\ge 1} n P_n(\theta)/\sum_{n\ge 1} P_n(\theta) 
\sim \theta^{\alpha/\beta}$.
The data colapse of $P_n(t)$ is shown in Fig.\ref{fig4_GM}. It was obtained 
for $\alpha = 1.47 \pm 0.05$ and $\beta = 1.23 \pm 0.05$.

\epsfigs{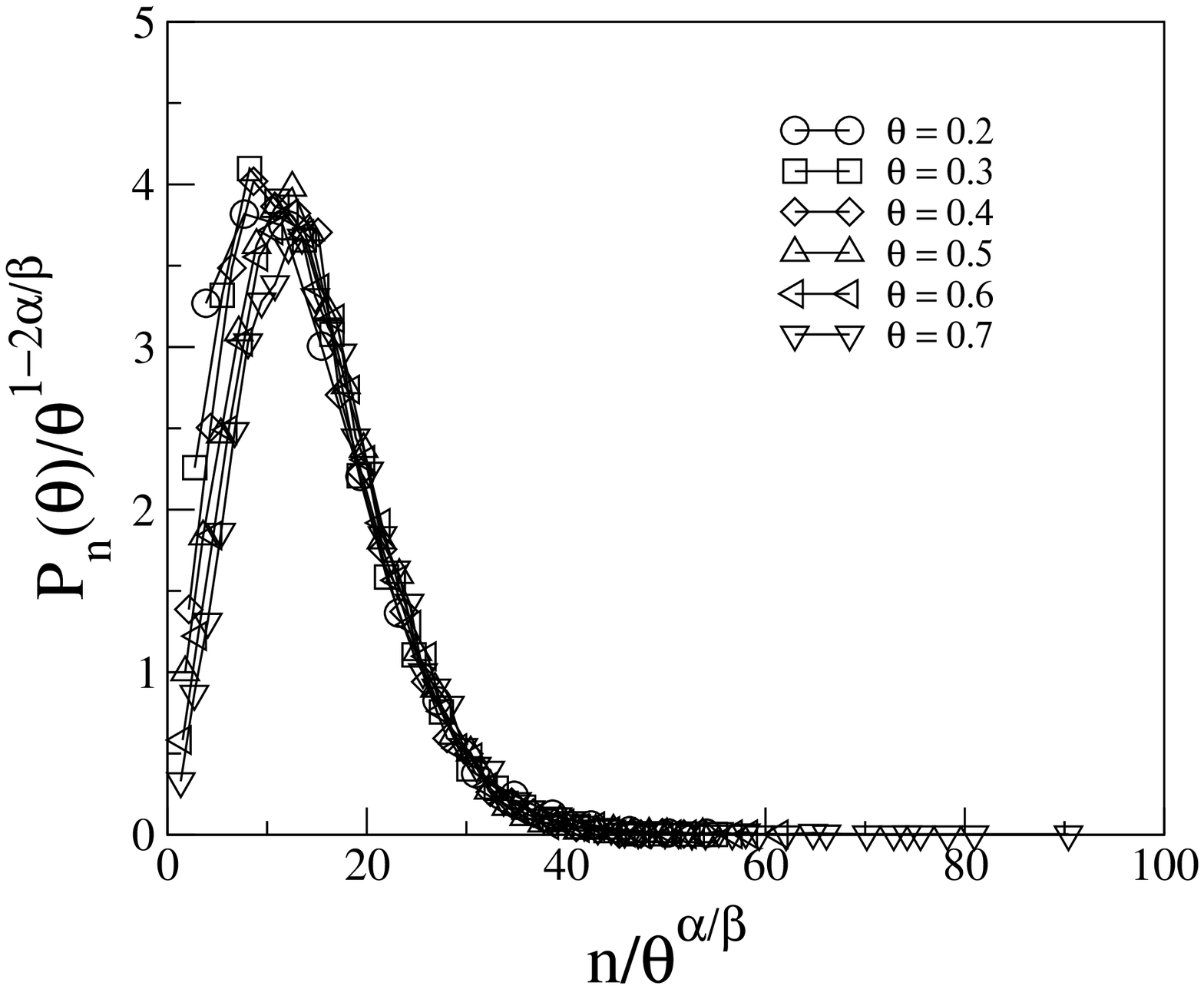}{Scaling of the mass distribution function Pn(t) with the 
system coverage. The values of $\alpha$ and $\beta$ are respectively, 1.23 
and 1.47 with an error of $0.05$ in both cases. 
($T=500K$ and $D=200$)}{fig4_GM}

\epsfigs{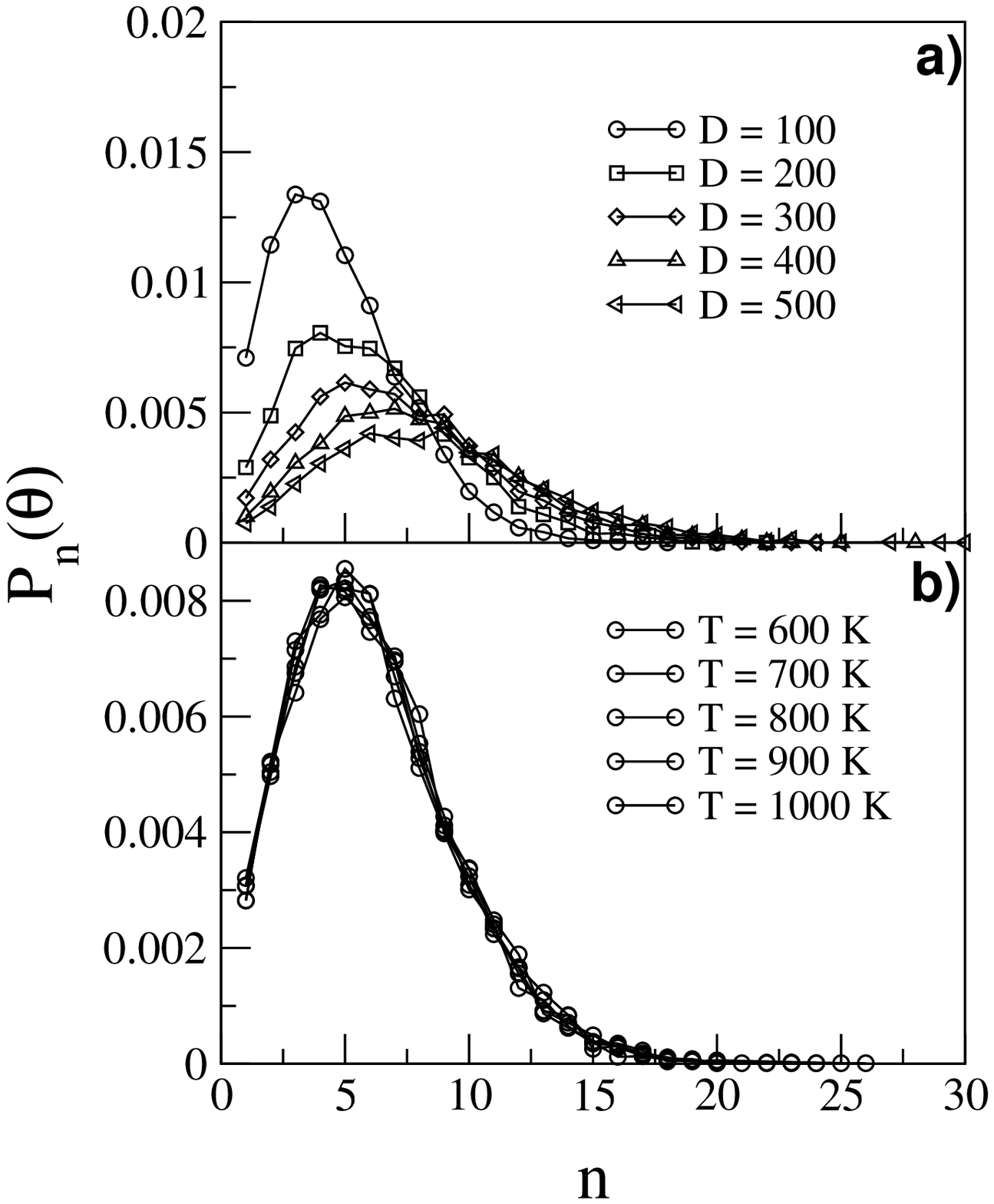}{Variations in the mass distribution when the number of 
diffusion steps 
$D$ and the temperature $T$ changes. a) With the increase 
in $D$, the mass distribution keeps the same basic shape, but it's width 
increases while it's height decreases in a form similar to the one observed 
in Fig.\ref{fig1_GM} when we increased the coverage of the 
system ($T=500$ and $\theta=0.3$). The 
temperature doesn't cause any important changes in the form that the 
system organizes itself.($T=500K$, $\theta=0.3$)}{fig2_GM}

When we keep the value of the coverage fixed and vary $D$, the 
mass distribution function behaves in a manner similar to the one 
described above. As $D$ increases, the distribution function widens 
and flattens. This is due to the fact that, with a larger number of 
diffusion steps, the adatom has a larger probability of diffusing 
away from the local minimum of energy and coalescing with other particles 
present in the system, originating larger clusters.
On the other hand, when we change the temperature, nothing seems to happen, 
the distribution function maintains its basic properties.

\epsfigs{fig4_GM}{Correlation function at various coverages. In a) we observe 
the evolution of the correlation function as the coverage increases. It is 
clearly seen that the position of the minimum tends to move to larger values
 of r ($T=500K$ and $D=200$). b) A similar precession of the minimum is observed when $D$ is increased ($\theta=0.3$ and $T=500$).
c) The correlation function doesn't seem to be affected by the changes in T. 
This is probably due to the fact that only the diffusing adatom feels the 
effect of the temperature ($D=200$ and $\theta=0.3$).}{fig3_GM}

As it can be seen in Fig\ref{fig3_GM} a, the correlation function $C(r,t)$, 
defined in Eq.\ref{corr}, displays a characteristic behavior, starting 
at values of the order of 0.2 and decreasing until a minimum, at distances 
of the order of ten lattice units, and finally oscillating with decreasing 
amplitude around zero until it becomes effectively zero at distances of 
the order of 40. 
As it is easily seen from Fig.\ref{fig3_GM}, the correlation function 
always maintains the same shape, even when $\theta$ or $D$ are increased. 
In the latter case, the position of the minimum seems to move to larger 
distances as $D$ is increased, which means that the system becomes more 
correlated. Once again, the temperature doesn't have any real effect on 
the results.

A parameter that usually has a great importance in experimental study 
of this type of 
systems is the temperature. As can be seen in the previous figures, the 
temperature doesn't seem to have a great influence in the final result, 
contrary to what was expected. This peculiar behavior of the 
system can probably be explained by the 
fact that only the adatom that is being currently absorbed feels the 
temperature, the rest of the system being actually frozen. If we allowed 
the system to rearrange itself after the deposition of each particle, 
the temperature dependence would probably be more realistic, but the 
computation time required would also be much higher.

\section{Conclusion}
In conclusion, the simulations carried in a one dimensional system 
in the submonolayer regime with long range interactions allowed us to 
observe the mechanism of self-organization through the formation of islands 
of similar size over all system. The influence of different factors 
on this behavior was tested. Surprisingly, it does not show any 
dependence on the temperature, at least for the tested range 
($300 K<T<1000 K$). 

Ordering occurs to minimize the repulsive elastic interactions between 
absorbed atoms. This self-organization breaks down when the 
coverage gets large which makes the adatom have less space to find 
an equilibrium position and makes the coalescence events become more and 
more frequent and finally dominate the dynamics of the system.
We are now extending this results to the $2D$ case. 

\section{Acknowledgments}
It is a pleasure to thank Serguei Dorogovtsev for collaboration 
in the early stage of this work. We also thank M.P. Santos for a 
critical reading of the manuscript. This work was funded in part 
by the project POCTI/1999/FIS/33141 (FCT-Portugal).

\vspace{20pt}

\noindent

$^\ast$    Electronic address: bgoncalves@breathe.com\\
$^\dagger$ Electronic address: jfmendes@fc.up.pt


\begin{references}
\bibitem{nota1} This method of implementation seems to be somewhat 
unphysical, but gives the same qualitative results with less computational 
effort that the usual method of allowing the particle to diffuse from the 
begining. This is due to the fact that the adatom will always tend to 
difuse to the local energy minimum. Our method only acellerates this 
process without changing the physics behind it.
\end{references}
\end{document}